\documentclass[11pt,onecolumn,amssymb,nofootinbib]{revtex4}
\usepackage{amsmath, amsthm, amscd, amssymb}
\usepackage{bm}
\usepackage{bbm}

\begin{document}

\title{\bf Heating Through Phonon Excitation Implied by Collapse Models}

\author{Stephen L. Adler}
\email{adler@ias.edu} \affiliation{Institute for Advanced Study,
Einstein Drive, Princeton, NJ 08540, USA.}

\begin{abstract}
We calculate the rate of  heating through phonon excitation implied by the noise postulated in mass-proportional-coupled collapse models, for a general noise power spectrum.  For white noise with reduction rate $\lambda$, the phonon heating rate reduces to the standard formula, but for non-white noise with power spectrum $\lambda(\omega)$,
the rate $\lambda$ is replaced by $\lambda_{\rm eff}=\frac{2}{3 \pi^{3/2}} \int d^3w e^{-\vec w^2} \vec w^2 \lambda(\omega_L(\vec w/r_c))$,
 with $\omega_L(\vec q)$ the  longitudinal acoustic phonon frequency as a function of wave number $\vec q$, and with $r_C$ the noise
correlation length.  Hence if the noise power spectrum is cut off below $\omega_L(|\vec q| \sim r_c^{-1})$,  the heating rate is sharply reduced.
\end{abstract}
\maketitle

There is increasing interest in testing wave function collapse models \cite{collapse},  by searching for effects associated with the noise which drives wave function collapse when nonlinearly coupled in the Schr\"odinger
equation.  A recent
cantilever experiment of Vinante et al. \cite{vin}  has set noise bounds consistent with the enhanced noise strength  \cite{adler1} needed to make latent image formation a trigger for state vector collapse, and reports a possible noise signal. Various other suggested experiments \cite{alternative} focus on noise-induced motions or heating
of small masses or collections of oscillators, assuming a white noise spectrum. Since recent experiments on gamma ray emission from germanium \cite{ge} have
shown that with the enhanced noise strength of \cite{adler1}, a white noise spectrum is experimentally ruled out, it becomes important to take the effects of a cutoff in the noise
spectrum into account.  In this paper we focus on noise-induced heating, motivated by the astute  observation of Vinante \cite{vin1} that since the noise wave number density
is peaked near $|\vec q|\sim r_c^{-1}$,  heating effects will be reduced if the noise spectrum cuts off below the longitudinal acoustic phone frequency associated with
the wave number peak.  Our aim is to give a quantitative calculation of this effect; its application to possible experiments involving bulk heating effects will be given
elsewhere \cite{adlvin}

Consider a system in initial state $i$ with energy $E_i= \hbar \omega_i$ at time $t=0$, acted on by a perturbation $V$ which at time $t$ leads to a transition to
a state $f$ with energy $E_f=\hbar \omega_f$.  Working in the interaction picture, the transition amplitude $c_{fi}(t)$ is given by
\begin{equation}\label{tranamp}
c_{fi}(t)= -\frac{i}{\hbar}\int_0^t V_{fi}(t^{\prime})e^{i\omega_{fi}t^{\prime}}dt^{\prime}~~~,
\end{equation}
with $\omega_{fi}=\omega_f-\omega_i$.  For $V$ we take the noise coupling in the mass-proportional continuous spontaneous localization (CSL) model,
\begin{align}\label{noise}
V=&\int d^3z \frac{dW_t(\vec z)}{dt} {\cal V}(\vec z,\{\vec x\}) ~~~,\cr
{\cal V}(\vec z,\{\vec x\})=&-\frac{\hbar}{m_N} \sum_{\ell} m_{\ell} g(\vec z-\vec x_{\ell} )~~~,
\end{align}
where we have followed the notation used in \cite{adlerram}.  Here $\vec x_{\ell} $ are the coordinates of atoms of mass $m_{\ell}$, $g(\vec x)$ is a spatial correlation function, conventionally taken as a Gaussian
\begin{equation}\label{correl}
g(\vec x)=(2\pi)^{-3/2}\, (r_c)^{-3} e^{-{\vec x}^2/(2 r_c^2)}=(2\pi)^{-3} \int d^3q e^{-r_c^2 \vec q^{\,2}/2\,-i\vec q \cdot \vec x}~~~,
\end{equation}
and the non-white noise has expectation ${\cal E}$
\begin{equation}\label{noiseex}
{\cal E}\left[\frac{dW_t(\vec x)}{dt}\, \frac{dW_{t^{\prime}}(\vec y)} {dt^{\prime}}\right] =\frac{1}{2\pi} \int_{-\infty}^{\infty} d\omega
\gamma(\omega) e^{-i\omega(t-t^{\prime})}\delta^3(\vec x-\vec y)~~~,
\end{equation}
with $\gamma(\omega)=\gamma(-\omega)$ related to the reduction rate parameter $\lambda(\omega)$ by
\begin{equation}\label{gamal}
\gamma(\omega)=8\pi^{3/2} r_c^3\lambda(\omega)~~~.
\end{equation}

We wish now to calclulate the expectation ${\cal E}[E(t)]$ of the energy attained by the system at time $t$, given by
\begin{equation}\label{enex}
{\cal E}[E(t)]={\cal E}[\sum_f \hbar \omega_{fi}|c_{fi}(t)|^2]~~~.
\end{equation}
Substituting Eqs. \eqref{tranamp} -- \eqref{gamal}, carrying out integrations, and using the formulas \cite{cohen}
\begin{align}\label{deltaform}
\int_0^t dt^{\prime} e^{i(\omega_{fi}-\omega) t^{\prime}} =& \frac{e^{i(\omega_{fi}-\omega) t}-1}{i (\omega_{fi}-\omega)}
\equiv 2\pi  e^{i(\omega_{fi}-\omega) t/2}\delta^{(t)}(\omega_{fi}-\omega)~~~,\cr
[\delta^{(t)}(\omega_{fi}-\omega)]^2\simeq &\frac{t}{2\pi} \delta^{(t)}(\omega_{fi}-\omega)~~~,\cr
\end{align}
we find in the large $t$ limit the formula for the energy gain rate
\begin{equation}\label{rate}
t^{-1} {\cal E}[E(t)]=\frac{r_c^3}{\pi^{3/2}m_N^2} \int d^3 q \sum_f e^{-r_c^2 \vec q^2} \lambda(\omega_{fi})\hbar\omega_{fi}
\left|\left(\sum_{\ell} m_{\ell} e^{i\vec q \cdot \vec x_{\ell} }\right)_{fi}\right|^2~~~.
\end{equation}

The next step is to evaluate the matrix element appearing in Eq. \eqref{rate} by introducing  phonon physics, following the exposition in the text of Callaway \cite{cal}.  We
consider first the simplest case of a monatomic lattice with all $m_{\ell} $ equal to $m_A$, independent of the index $\ell$, and write the atom coordinate $\vec x_{\ell}$ as
\begin{equation}\label{coord}
\vec x_{\ell}=\vec R_{\ell}+\vec u_{\ell}~~~,
\end{equation}
with $\vec R_{\ell}$ the equilibrium lattice coordinate and with $\vec u_{\ell}$ the lattice displacement induced by the noise perturbation.   Writing
\begin{equation}\label{internsum}
\sum_{\ell} m_{\ell} e^{i\vec q \cdot \vec x_{\ell} }=m_A \sum_{\ell} e^{i\vec q \cdot  \vec R_{\ell}}e^{i\vec q \cdot  \vec u_{\ell} }~~~,
\end{equation}
we note that since the Gaussian in Eq. \eqref{rate} restricts the magnitude of $\vec q$ to be less than of order of $r_c^{-1}$, with $r_c \sim 10^{-5} {\rm cm}$, whereas
the magnitude of the lattice displacement is much smaller than $10^{-8} {\rm cm}$, the exponent in $e^{i\vec q \cdot  \vec u_{\ell} }$ is a very small quantity.
So we can Taylor expand to write
\begin{equation}\label{expan}
e^{i\vec q \cdot  \vec u_{\ell} }\simeq 1+ i\vec q \cdot  \vec u_{\ell}~~~.
\end{equation}
The leading term $1$ does not contribute to energy-changing transitions, so we have reduced the matrix element in Eq. \eqref{rate} to the simpler form
\begin{equation}\label{matrix1}
\left(\sum_{\ell} m_{\ell} e^{i\vec q \cdot \vec x_{\ell} }\right)_{fi}\simeq im_A \left( \sum_{\ell} e^{i\vec q \cdot  \vec R_{\ell}} \vec q \cdot \vec u_{\ell}\right)_{fi}
~~~,~~~~f \neq i~~~.
\end{equation}
The approximation leading to Eq. \eqref{matrix1}  is a phonon analog of the electric dipole approximation made in electromagnetic radiation rate calculations.

We now substitute the expression \cite{cal} for the lattice displacement in terms of phonon creation and annihilation operators,
\begin{equation}\label{disp}
\vec u_{\ell}= \frac{\Omega}{8\pi^3} \left(\frac{\hbar {\cal N}}{m_A}\right)^{1/2} \sum_j \int \frac{d^3 k}{(2\omega_j(\vec k))^{1/2}}
\big[\vec e^{\,(j)}(\vec k)   e^{i\vec k \cdot \vec R_{\ell}} a_j(\vec k)+ \vec e^{\,(j)*}(\vec k) e^{-i\vec k \cdot \vec R_{\ell}} a_j^{\dagger}(\vec k)\big]~~~,
\end{equation}
where the sum on $j$ runs over the acoustic phonon polarization states, and where $\Omega$ and ${\cal N}$ are respectively the lattice unit cell volume, and
 the number of unit cells.  Taking the initial state $i$ to be the zero phonon state, only the $a_j^{\dagger}$ term
in Eq. \eqref{disp} contributes, and we can evaluate the sum over lattice sites $\ell$ in Eq. \eqref{matrix1} using the formula \cite{cal}
\begin{equation}\label{sum1}
\sum_{\ell} e^{i (\vec q-\vec k)\cdot \vec R_{\ell}}= \frac{8\pi^3}{\Omega}\delta^3(\vec q-\vec k)~~~.
\end{equation}
Carrying out the $\vec k$ integration, noting that $\vec q \cdot \vec e^{\,(j)}(\vec q)$ selects the longitudinal phonon with frequency $\omega_L(\vec q)$, defining $\vec w=r_c \vec q$, writing $M={\cal N} m_A$ for the total system mass, and assembling all the pieces, we arrive at the answer
\begin{align}\label{final1}
t^{-1} {\cal E}[E(t)]=&\frac{\hbar^2 M}{m_N^2 r_c^2} \frac{1}{2 \pi^{3/2}} \int d^3w e^{-\vec w^2} \vec w^2 \lambda(\omega_L(\vec w/r_c))=\frac{3}{4} \frac{\hbar^2\lambda_{\rm eff}  M}{m_N^2 r_c^2}~~~,\cr
\lambda_{\rm eff}\equiv &\frac{2}{3\pi^{3/2}} \int d^3w e^{-\vec w^2} \vec w^2 \lambda(\omega_L(\vec w/r_c))~~~.\cr
\end{align}
In the white noise case, where  $\lambda(\omega)$ is a constant $\lambda$, we can pull it outside the $\vec w$ integral and use
\begin{equation}\label{pulled}
\int d^3 w e^{-\vec w^2} \vec w^2 =\frac{3}{2} \pi^{3/2}
\end{equation}
to get the standard formula \cite{gain}
\begin{equation}\label{stdform}
t^{-1} {\cal E}[E(t)] = \frac{3}{4} \frac{\hbar^2 \lambda M}{m_N^2 r_c^2}~~~.
\end{equation}
When the noise spectrum has a cutoff below $\omega_L(\vec q)$ for $|\vec q| \sim r_c^{-1}$, the energy gain rate is sharply reduced.

Although we have derived the result of Eq. \eqref{final1} for the case of a monatomic lattice and a zero phonon initial state, the result is more general.
For a multi-atom unit cell, the same answer holds, with $m_A$ the sum of masses in the unit cell, and with $\omega_L(\vec q)$ again the longitudinal acoustic
phonon frequency. In the multi-atom case  the formula of Eq. \eqref{final1} neglects optical phonon contributions, but these are the ``internal excitations'' that are neglected in
the derivation of the center-of-mass energy gain formula of Eq. \eqref{stdform}. When the initial state is constructed from  $n$-phonon states,
as in a thermal ground state, the $a^{\dagger}$ term in Eq. \eqref{disp} contributes a term proportional to $(n+1)\omega_L$ to the
energy gain, while the $a$ term in Eq. \eqref{disp} contributes a corresponding term proportional to $-n \omega_L$ to the energy gain; the sum of the two terms
is proportional to $(n+1-n) \omega_L = \omega_L$, so $n$ drops out and the formula of Eq. \eqref{final1} is recovered.  This simplification  could have been anticipated
from our earlier analysis of the noise-induced energy gain by an oscillator \cite{adlerosc}, which showed that the rate of energy gain is a constant independent of the number of
oscillator quanta that are present.

I wish Andrea Vinante for an email that stimulated this paper, and to thank Angelo Bassi for helpful conversations.

\bigskip
{\bf Added Note}

Apart from updating Ref. [7], the preceding body of this paper is identical to the version posted on arXiv on Jan. 1, 2018.  Andrea Vinante has called our attention to a paper by M. Bahrami \cite{bahrami} posted on Jan. 11, with an update on Jan. 14, in which a similar calculation is done.
For a monatomic lattice, Bahrami's result and ours are in agreement.  In his Jan. 14 posting,
Bahrami gives a formula for the case of a multi-atom unit cell, which he notes disagrees with our statement that this gives the same result as the monatomic case.  Bahrami's
multi-atom formula is incorrect, as a result of his using the wrong normalization for
the phonon polarization vectors, and does not reduce to the standard formula in the white noise case when $\lambda(\omega)$ is a constant $\lambda$.
In this version of our paper, we have added an Appendix giving a brief derivation of
the correct result in the multi-atom case.

\bigskip
{\bf Later Added Note}  

Bahrami agrees, and will revise his posting.

\bigskip
{\bf Appendix: Brief derivation of the formula for the multi-atom case}

In the monatomic case, focusing  only on the atomic mass factors and longitudinal phonon polarization vectors, Eqs. \eqref{matrix1} and \eqref{disp} give a factor
\begin{equation}\label{mono}
m_A^{1/2} \vec{e}^{\,(L)*}(\vec k)\simeq m_A^{1/2} \vec{e}^{\,(L)*}(\vec 0)~~~.
\end{equation}
After the $\simeq$ sign we have used the fact, noted after Eq. \eqref{internsum},
that the correlation length $r_C$ allows only contributions from phonon wavelengths that are long on a lattice scale, corresponding to $\vec k \simeq \vec 0$.  In the multi-atom case, focusing only on acoustic phonons,\footnote{Optical phonons leave the unit cell
center of mass stationary, so obey $\sum_{\kappa}m_{\kappa}^{1/2} \vec{e}_{\kappa}^{\,(s)}(\vec 0) =0$ for any optical phonon mode $s$. Hence for mass-proportional
noise coupling, optical phonons do not contribute to the energy gain rate to leading
order in $a/r_C$, with $a$ the unit cell dimension.}  the left-hand side of Eq. \eqref{mono} is replaced by
\begin{equation}\label{multi}
m_{\kappa}^{1/2} \vec{e}_{\kappa}^{\,(L)*}(\vec k)~~~,
\end{equation}
corresponding to Eqs. (1.4.22a,b) of  \cite{cal}, with $\kappa$ labeling an atom in the
multi-atom unit cell.  Referring now to the unnumbered equation in Callaway \cite{cal} between his  Eqs. (1.1.22) and (1.1.23), which we write \big(using the fact that for $\vec k=0$ the polarization vectors are real numbers; see Callaway Eq. (1.1.21)\big) as
\begin{equation}\label{polvec}
m_{\kappa}^{-1/2} \vec{e}_{\kappa}^{\,(L)*}(\vec 0)=m_{\kappa}^{-1/2} \vec{e}_{\kappa}^{\,(L)}(\vec 0)=\vec C~~~,
\end{equation}
with $\vec C$ a constant,
we see that the longitudinal polarization vectors are no longer unit normalized, as
in the monatomic case.  Instead, the normalization is given in Eq. (1.1.18a) of
\cite{cal},
\begin{equation}
\sum_{\kappa} \vec{e}_{\kappa}^{\,(L)*}(\vec 0) \cdot \vec{e}_{\kappa}^{\,(L)}(\vec 0)
=1~~~,
\end{equation}
which on substituting Eq. \eqref{polvec} gives
\begin{equation}\label{cmag}
|\vec C|= \left(\sum_{\kappa} m_{\kappa}\right)^{-1/2}~~~,
\end{equation}
and implies for small $\vec k$
\begin{equation}\label{polvec1}
m_{\kappa}^{1/2}\hat k \cdot  \vec{e}_{\kappa}^{\,(L)*}(\vec k) \simeq  m_{\kappa}  |\vec C| =  m_{\kappa} \left(\sum_{\kappa} m_{\kappa}\right)^{-1/2}~~~.
\end{equation}
Recalling Eqs. \eqref{expan}--\eqref{sum1}, summing over $\kappa$ to get the total contribution to the one-phonon creation amplitude, we have
\begin{equation}\label{summing}
 \sum_{\kappa} m_{\kappa} \left(\sum_{\kappa} m_{\kappa}\right)^{-1/2}~~~,
\end{equation}
which when squared gives a factor
\begin{equation}\label{squared}
\sum_{\kappa} m_{\kappa}=m_{\rm cell}~~~,
\end{equation}
which is the total atomic mass in the unit cell.   Thus the only change from
the monatomic to the multi-atomic case is the replacement of $m_A$ by $m_{\rm cell}$,
and since ${\cal N}m_{\rm cell}=M$, the total system mass, the monatomic formula
of Eq. \eqref{final1} is unchanged.  Heuristically, the reason for this is that, as
emphasized by Callaway, for $\vec k=0$ acoustic phonons Eq. \eqref{polvec} implies that
all ``...particles in each unit cell move in parallel with equal amplitudes'', and so
behave as a single particle with mass $m_{\rm cell}$.

\end{document}